\documentclass[12pt]{article}
\usepackage{epsf}
\def\beq{\begin{equation}}   \def\eeq{\end{equation}}

\begin{document}
\begin{titlepage}
\begin{flushright}
TPI-MINN-99/58-T  \\
UMN-TH-1830/99   \\
\end{flushright}

\vspace{0.6cm}

\begin{center}
\Large{{\bf Calculating the tension of domain wall junctions and vortices
in generalized Wess-Zumino models.}}

\vspace{1cm}

M. Shifman and T. ter Veldhuis
\end{center}
\vspace{0.3cm}

\begin{center}
{\em Theoretical Physics Institute, Univ. of Minnesota, 
Minneapolis,
       MN 55455}
\end{center}
\vspace{0.5cm}

\begin{abstract}
We study BPS saturated objects with axial geometry 
(wall junctions, vortices) in 
generalized Wess-Zumino models. It is observed that
the tension of such objects is negative in general
(although ``exceptional'' models are possible).
We show how an ambiguity in the definition of central
charges does not affect physical quantities, and
we comment on the stability of the junctions and
vortices. We illustrate these issues in two classes
of models with $Z_N$ symmetry. On the basis of
analytical large $N$ calculations and numerical calculations
at finite $N$, we argue that the domain wall junctions
in these models are indeed BPS saturated, and we
calculate the junction tensions explicitly.
\end{abstract}

\vspace{0.5cm}

\begin{flushleft}
E-mail: shifman@physics.spa.umn.edu, veldhuis@hep.umn.edu
\end{flushleft}
\end{titlepage}

\section{Introduction}

Studying the BPS domain walls  in supersymmetric
theories  
is interesting, especially in gauge theories \cite{DS2}, because 
one can get information on 
non-trivial dynamical features. For example, it was found 
 that the tension of such domain
walls determines the decay rate of \lq\lq false \rq\rq vacua in large $N$
supersymmetric gluodynamics \cite{W,S}.  

It was noted \cite{AT}
that theories with either
a $U(1)$ or $Z_N$ global
symmetry may contain BPS objects with
axial geometry.
Vortex-like objects which conserve
$1/4$ of the original supersymmetry were considered
in \cite{CS}, while domain wall junctions of the
hub and spoke type, which also preserve $1/4$ of supersymmetry, were 
discussed in \cite{AT,GT,CHT}. General considerations of the
tensorial central charges
responsible for the saturation of
the vortices and junctions were presented in \cite{AGIT,FP,GS}.
The existence of BPS wall junctions also
leads to non-trivial dynamical consequences. For instance,
in \cite{GaS} the large $N$ behavior of the
BPS junction tension was used to argue in favor of soliton-like
heavy hadrons with mass proportional to $N$ in supersymmetric
gluodynamics.

The study of 
possible dynamical implementation of the 
BPS wall junctions has just begun.
Apart from the original works in which 
the existence of such junctions was first noted,
they were discussed in \cite{GaS,B,OINS,BT}. 
In \cite{GaS} a generalized Wess-Zumino model, which we will
refer to as Model I,
with one chiral
superfield $X$ and the superpotential
\begin{equation}
{\cal W} = N\left\{ X + \frac{N}{N+1}\left(\frac{X}{N}\right)^{N+1}\right\}
\,,
\label{modone}
\end{equation}
was considered. 
This model
is related to the Veneziano-Yankielowicz \cite{VY} effective
Lagrangian, which, in turn, models supersymmetric 
Yang-Mills theory. The parameter $N$ is an integer that corresponds to the
the number of colors of the
gauge group ${\rm SU}(N)$ of SUSY gluodynamics. The superpotential (1) was
originally suggested for that purpose in \cite{DGK}. At any given $N$ it 
possesses a $Z_N$ symmetry which is spontaneously broken. In the large
$N$ limit an analytic solution for the BPS domain walls
was found in \cite{DK}. The most essential feature of the
solution is the fact that the wall width scales as
$N^{-1}$, and vanishes at $N\rightarrow \infty$. This result, in
conjunction with two BPS junction configurations presented
in \cite{GaS} at $N=\infty$, implies that
BPS junctions do exist at large $N$ in the theory with
superpotential (1). The junction tension was found to scale
as $N^2$ at large $N$.

In \cite{OINS} the first analytic solution for a BPS junction was found
in a specific generalized Wess-Zumino model. Among
stimulating findings in this work
is the fact that the junction tension turned out to be
negative in this model.
The model has $Z_3$ symmetry. It is derived from a
${\rm SU}(2)$ Yang-Mills theory with 
extended supersymmetry (${\cal{N}} = 2$) and one matter
flavor perturbed by an adjoint scalar mass.
The original model contains three pairs of chiral superfields
and, in addition, one extra chiral superfield.
In fact, for the purpose of studying the
BPS walls and wall junctions, the
model can be further simplified. It can be easily 
stripped of its inessential features and cast into the
form of a $Z_N$ symmetric model. ($N$ is an arbitrary
integer, not necessarily equal to $3$). The generalized
and simplified model, which we refer to as Model II, contains $N$ chiral
superfields $M_k$ with $k=1,2,...,N$ and another chiral
superfield $T$. The superpotential is

\beq
{\cal W} = N\left\{-A\left[\sum_{k=1}^{N}\left(T-Nq^{k-1}e^{-\pi i/N}\right)M_k^2\right]
+BT\right\}\,,
\label{modtwo}
\eeq
where $q=e^{i 2 \pi/N}$. The constants $A$ and $B$ are assumed to be $N$ 
independent and positive.

Model I is invariant under the transformation
\beq
X(x,\theta )\to e^{2\pi i k/N} X(x,\theta ')\,,\quad \theta ' = 
e^{-\pi i k/N}\theta\, ;
\eeq
the corresponding transformation of the superpotential is
\beq
{\cal W}(X(x,\theta )) \to e^{2\pi i k/N}{\cal W}(X(x,\theta '))\,.
\eeq
Model II is invariant under the transformation
\beq
T(x,\theta )\to e^{2\pi i k/N} T(x,\theta ')\,,\quad 
M_l(x,\theta )\to M_{l-k} (x, \theta ')\,\quad
\theta ' = e^{-\pi i k/N}\theta\, ;\label{symmII}
\eeq
the corresponding transformation of the superpotential is
\beq
{\cal W}(T(x,\theta ),M(x,\theta)) \to 
e^{2\pi i k/N}{\cal W}(X(x,\theta '), M(x,\theta '))\,.
\eeq

Model I is among the simplest supersymmetric models allowing
one to study BPS junctions and calculate tensions.
\footnote{It is worth noticing that in a different context its
$2$-dimensional reduction was studied in \cite{FMV}, where
it was shown to be integrable.} The model has $N$ distinct vacua in
which the scalar component of $X$ takes the vacuum
expectation value $N e^{i  \pi (2 k - 3)/N}$, with $k=1,2,..,N$.
The perturbative spectrum consists of a chiral
multiplet with mass $m=N$.

Let us now briefly discuss the vacuum structure
and spectrum of Model II. In this model there are $2N$ distinct,
physically equivalent
vacua. In these vacua the scalar components of the $T$ and $M$ fields
take the vacuum expectation values
\beq
T=N e^{i \pi (2k-3)/N}\,, \quad M_k=\pm \sqrt{\frac{B}{A}}\,,
\quad M_l=0\,\quad {\rm for}\,\quad l\neq k\,, \label{vacII}
\eeq
with $k=1,2,..,N$. The perturbative spectrum contains
two chiral multiplets, linear combinations of the field $T$
and the field $M_k$ that obtains a vacuum expectation value,
with mass $m_{1,2}=2 \sqrt{AB} N$, and $(N-1)$ chiral
multiplets with mass
\beq
m_k= 4 N^2 A \left( 1-\cos \pi(k-1)/N \right),
\eeq
for $k=2,3,..,N$.

Apart from the $Z_N$ symmetry, Eq.(\ref{symmII}), Model II has 
additional $Z_2^N$ symmetry, since
any of the fields $M_k$ can be transformed as
$M_k \rightarrow -M_k$. The superpotential in Eq.(\ref{modtwo})
is obviously invariant under this transformation. In our consideration
of the wall junctions below we will disregard the $Z_2^N$
symmetry, limiting ourselves just to a $Z_N$ family of
vacua. (This corresponds to the choice of the $+$ sign in
front of $\sqrt{A/B}$ in Eq.(\ref{vacII})). The walls and junctions 
of the type where (some of)
the vacua lie outside this family may or may not
be BPS saturated. We leave this issue open for future
investigations.

Needless to say that in both models I and II the
walls connecting the distinct vacua and the corresponding
junctions are topologically stable. (The issue of stability
of the BPS  junctions is discussed separately
in Sect.(\ref{stability}).) They need not be BPS saturated, generally
speaking. We know, empirically, that sometimes in certain models
equations of the BPS saturation have no
solutions with the proper boundary conditions \cite{FMV,SV}
even in the case of isolated walls, let alone the wall
junctions.

It was known
previously that the isolated walls in  Model I are
saturated, see \cite{DGK}. In Appendix A we give an
analytical solution to the BPS equation for the
domain walls in Model II for a specific ratio of
the parameters $A$ and $B$;  we have verified
numerically that these walls also saturate the
BPS bounds for other values of the parameters.
In Sect.(\ref{saturation}) we provide evidence that
the domain wall junctions in Models I and II are
also BPS saturated.
Were the walls and junctions 
non-saturated, one could still calculate the relevant
central charges. They would then represent lower bounds
on the corresponding tensions.

Most considerations in this paper concern the basic domain
wall junction, the junction with $N$ sectors in a model with
$Z_N$ symmetry. But in Appendix B we present an analytic solution
to the BPS equation for the class of triple junctions in Model II for
a specific ratio of the parameters $A$ and $B$, and $N$ a multiple
of three. For $N=3$, the basic junction and the triple junction
coincide. Non-basic junctions in Model I are studied in \cite{BT}.
Networks of domain walls in Model I are studied in \cite{Sa,BB}.

In this paper we consider BPS wall junctions and calculate
their tensions in Models I and II specified above. We also 
consider general aspects, 
including the sign of the tension, the ambiguity
of the central charges, and stability,
that are valid beyond Model I and II.
These models are used to illustrate the general
assertions. 

The organization of our paper
is as follows. In Sect.(\ref{general}) we describe a general
method to calculate the BPS bound on the tension of
domain wall junctions. We show that the junction tension
is typically negative, although we do not exclude exotic
models with positive tension. We then use this
method to explicitly calculate the junction tension in Models I and II.
(Secs. 3 and 4).
For Model I we calculate the junction tension
analytically in next to leading order in $1/N$, and 
numerically for finite values of $N$. For Model II 
the junction tension is calculated analytically for
all values of $N$ and all values of the parameters $A$ and $B$.
In Sect.({\ref{ambiguity}) we show that ambiguities in the
central charges cancel in the calculation of the 
junction tension, and we illustrate this  cancellation by recalculating
the tension in Model II with a different definition of the
central charges. In Sect.({\ref{saturation}) 
we demonstrate by numerical -- and in certain limits,
analytical -- calculations that in both models I and II the
wall junctions are, in fact, saturated. 
In Sect.(\ref{stability}) we comment on the stability
of the  BPS junctions.
The appendices contain analytical results for domain walls and the
triple junction in Model II.

\section{Generalities of the solutions with the axial geometry \label{general}}

For both Models I and II it is possible to calculate, using the
general theory of the central charges, the BPS 
bounds on the junction tension. The general
formula worked out in \cite{GS} implies
\begin{equation}
\frac{M}{L} = - \oint a_k dx_k + 2 \oint dn_k S_k,
\label{junmasseq}
\end{equation}
where $a_k$ is an axial current. At the classical level only the component
of the current built from the scalar fields is important. Moreover,
\begin{equation}
S_{1,2} = \left\{ {\rm{Re}} \Sigma, {\rm{Im}} \Sigma \right\},
\end{equation}
where $\Sigma$ is related to the superpotential.
The integrals run over the large circle (see Fig.(\ref{configuration})) with $R
\rightarrow \infty$, and $d\vec{x}$ ($d\vec{n}$) is an infinitessimal vector
along (perpendicular to) the contour. The precise definitions of $a_k$ and
$\Sigma$ have (correlated) ambiguities. The sum in Eq.(\ref{junmasseq})
is unambiguous, however \cite{GS}. The above ambiguity can be exploited
to optimize the calculation
of the junction tension. This will be explained below in two particular
applications, Model I and II.

\begin{figure}
\epsfxsize=7cm
\centerline{\epsfbox{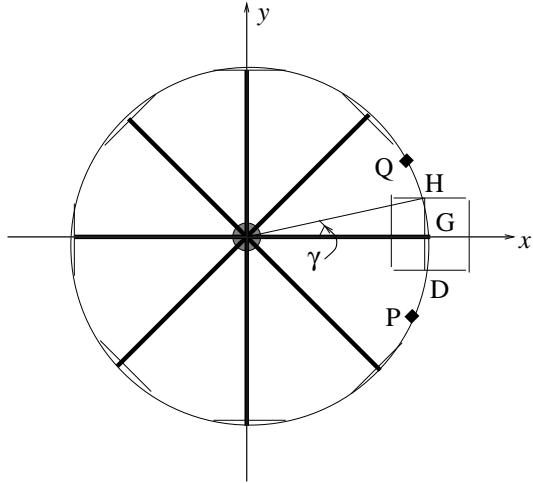}}
\caption{A ``standard'' convention regarding the wall junction.
The large circle of radius $R$ enters the definition of $E(R)$
while the circle with the segments removed is relevant to $\tilde E(R)$.
The box of size $L\times L$ around the point G is blown up in 
Fig.(\ref{detail}).
$L$ is an auxiliary parameter chosen to satisfy the constraint
$l\ll L \ll R$.}
\label{configuration}
\end{figure}

\begin{figure}
\epsfxsize=7cm
\centerline{\epsfbox{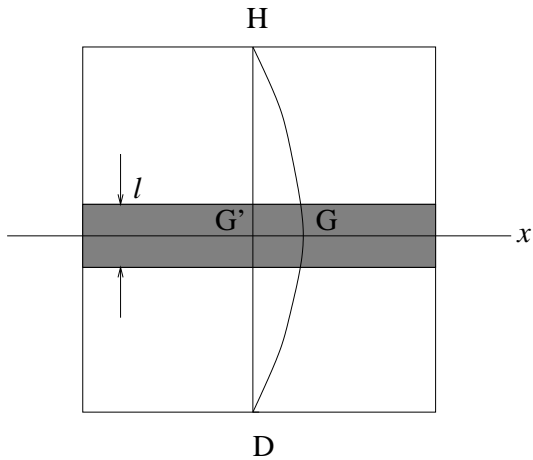}}
\caption{A blow up of the box in Fig.(\ref{configuration}). 
The shaded rectangle represents the wall,
the length of the interval GG' is of order $L^2/R$.}
\label{detail}
\end{figure}

We will consider the wall junction of the ``hub and spoke'' type.
The following conventions are convenient (although not necessary).
We will orient the wall spokes as indicated in Fig.(\ref{configuration}),
namely, the hub is at the origin, the ``first'' spoke
runs along the $\hat{x}$ axis in the positive direction, the ``second''
runs at an angle $2\pi/N$, and so on. In the point P
the theory ``sits'' in the first vacuum, in the point Q in the second, etc.
This configuration is topologically stable.

We will assume that the $Z_N$ symmetry is realized through
multiplication of (some of) the fields by a phase.
This is certainly true in Models I and II.
If on the junction solution (the lowest components of) the fields $X$, $T$ 
(generically, $\Phi$) rotate as
\beq
\Phi (\zeta e^{2\pi i k /N} ) = e^{2\pi i kr /N}\Phi (\zeta )\, ,
\eeq
we will say that the rotation weight of the field $\Phi$ is equal
to $r$.
Configurations with the positive rotation weight
will be referred to as junctions (vortices),
while those with the  negative weight will be referred to as
anti-junctions (anti-vortices).
It is worth reminding that
\beq
\zeta = x+i y\, , \quad \partial_\zeta = \frac{1}{2}(\partial_x - i \partial_y )\,.
\eeq
The BPS wall junctions satisfy the equations
\beq
2\partial_\zeta \Phi_k = e^{i\delta}\frac{\partial\bar{\cal W}}{\partial\bar\Phi_k}\,,
\label{eqjunc}
\eeq
where the phase $\delta$ depends on the phase of the superpotential (see below).
The equations for anti-junctions is
\beq
2\partial_{\bar\zeta} \Phi_k = e^{i\tilde\delta}
\frac{\partial\bar{\cal W}}{\partial\bar\Phi_k}\,,
\eeq
i.e. $\zeta$ is replaced by $\bar\zeta$. Thus, in our convention
the phase of $\Phi$ grows in the anticlock-wise directions for the
junction, and in the clock-wise direction for the anti-junction.

The superpotential can always be chosen in such a way
that in the vacuum P its phase is $\exp (-i\pi / N)$ while in the vacuum
Q it is $\exp (i\pi / N)$. See Fig.(\ref{configuration})
for the definition of P and Q.  (Note that ${\rm{Re}}\{ {\cal W} \}$ is then 
positive in both vacua). Then the phase $\delta$ in Eq. (\ref{eqjunc}) must 
be set to  zero in both Models I and II. 

For the wall junctions the rotation weights of $X$ and $T$ are obviously 
equal to $1$.

Now, we proceed to discuss  general features of the
tension associated with wall junctions.
In Fig.(\ref{configuration}) the energy of the junction configuration
(per unit length) is defined as the integral over the volume energy density 
over the area inside the circle, where it is assumed that the circle 
radius $R$ tends to infinity,
\beq
E(R) = \frac{M}{\rm length} = \int_{|\vec r |\leq R} {\cal H}(x,y)dxdy
=T_1 R + T_2 + ...\,,\quad R\to\infty\,.
\label{intdo}
\eeq
Here the dots denote terms vanishing in the limit $R\to\infty$.
It is tempting to say that $T_2$ is associated with the $(1/2,1/2)$ 
central charge while $T_1$ with the
$(1,0)$ central charge (see Eq.(\ref{junmasseq}), the first and
second term, respectively). We hasten to note that these central
charges, being considered individually, are
ambiguous \cite{GS}. It is only the combination
in Eq.(\ref{junmasseq}) which is fixed unambiguously.
It is intuitively clear that $T_1 = NT_{\rm wall}$ where $T_{\rm wall}$
is the tension of the isolated wall. Moreover, 
it is also clear that $T_2$
(which will be also referred to as $T_{\rm junction}$) is typically negative.
Indeed, from Eq.(\ref{intdo}) it follows that, for
small $R$, $E(R) = c R^2$, 
where $c$ is a numerical coefficient. 
At large $R$ the quadratic dependence on $R$ changes
to linear, see Eq.(\ref{intdo}). Matching these two scaling laws
straightforwardly, connecting the parabola and the
straight line at the value of R where their slope is equal, we conclude 
that $T_2$ is forced to be negative.
Models where it could be positive
would require, in essence,  two scales and an intermediate regime
between the parabolic and linear regime.
In order to illustrate the typical situation, we plot
the energy $E$ as a function of $R$ for Model II with $N=3$,
$B=\sqrt{3}$ and $A=B/9$ in Fig.(\ref{typicmod}). 

Actually,
this figure gives the energy $E$ 
defined as the integral over $ {\cal H}(x,y)$ over the  equilateral triangle
with distance $R$ from the center to each of the sides instead
of the energy in a circle. The orientation of the triangle is such that each
of the sides is perpendicular to a spoke. We made this  modification
because for this triangle the energy converges to the linear
relation Eq.(\ref{intdo}) exponentially fast, whereas for
a circle the departure from the linear behavior is of the
order $1/R$. (As explained below, the coefficient of the $1/R$ term is just a
geometrical factor taking into account the curvature of the circle
as it intersects the domain wall.)

 For the particular values of the parameters that
we chose, there is an analytical wall junction  solution of 
the BPS equations, which is presented in Appendix B. The energy
$E(R)$ can therefore be calculated exactly and is shown in
Fig.(\ref{typicmod}).
It is clear from Fig.(\ref{typicmod}) that even though there is a change
from a convex to a concave regime, $E(R)$ converges to the
linear regime so fast that $T_{\rm junction}$ is negative. 
The
tensions can be calculated analytically and take
the values $T_1=162$ (see Sect.(\ref{saturation})) and
$T_2=-27 \sqrt{3} / 2$ (see Sect.({\ref{tensII})). Similar
graphs for Model I junctions are presented in Ref.\cite{BT}.
For comparison,
we sketch what a similar graph would look like for an exotic
model with positive junction tension in Fig.(\ref{exoticmod}).

\begin{figure}
\epsfxsize=12cm
\centerline{\epsfbox{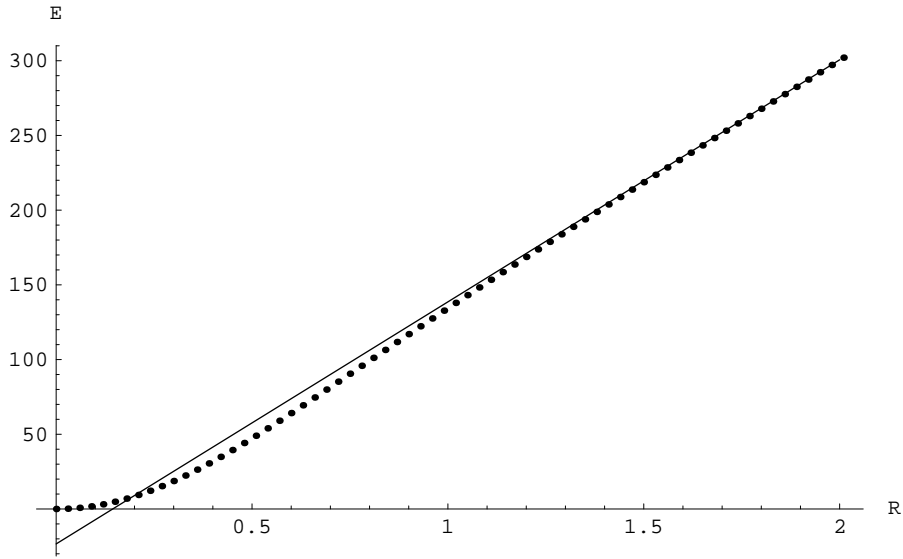}}
\caption{Energy $E$ as a function of radius $R$ (dotted line)
for the triple junction of model II with $N=3$, $A=B/N^2$ and $B=\sqrt{3}$, 
compared to the line $E= T_1 R + T_2$ (solid line). This example illustrates
the typical junction with negative tension. }
\label{typicmod}
\end{figure}
\begin{figure}
\epsfxsize=12cm
\centerline{\epsfbox{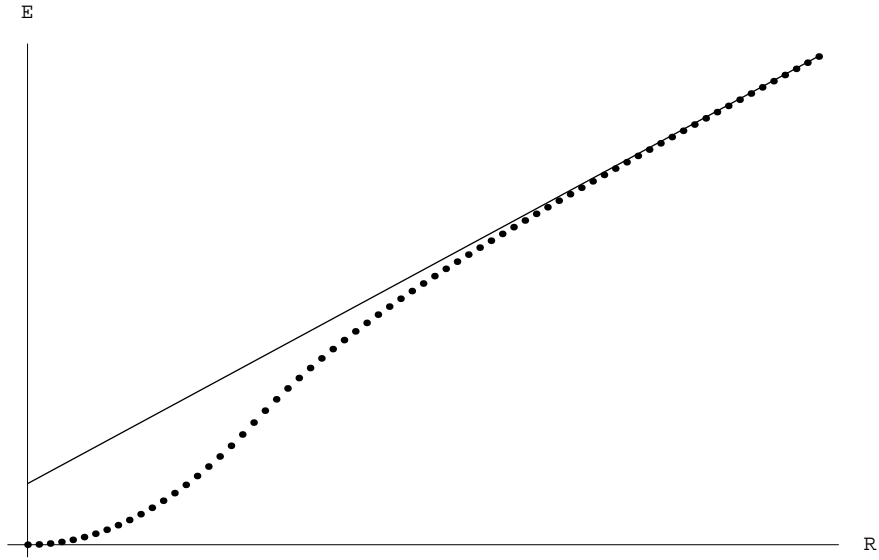}}
\caption{Sketch of the energy $E$ as a function of radius $R$
for a hypothetical, exotic model with the positive junction tension.}
\label{exoticmod}
\end{figure}

Below we
give a general prove of the fact that $T_1 = NT_{\rm wall}$,
and illustrate more quantitatively that $T_{\rm junction}$ is negative in
both Models I and II,  and in other ``natural'' models.
Of crucial importance is the fact that
the wall width (i.e. the transverse dimension inside which the
energy density is nonvanishing, while outside it
vanishes with exponential accuracy)
is finite (i.e. $R$ independent). This width is denoted by
$l$, see Figs. (\ref{configuration}) and (\ref{detail}).  
In both Models I and II, $l$ is actually $O(1/N)$.
At the moment we do not assume $N$ to be  large. It can be any integer
$> 2$. If $R\to\infty$, one can replace the integration domain in Eq. (\ref{intdo})
-- the interior of the circle can be replaced by the interior of
the circle with the segments removed. If $\tilde E(R)$ is defined as
the integral over the interior of the  circle with the segments removed, 
the difference
$\tilde E(R)-  E(R)$ tends to zero in the limit $R\to\infty$. (The length of 
the straight line DH is $L$,  where
$L$ is an arbitrary parameter, $l\ll L\ll R$.)
This can be clearly seen from Fig.(\ref{detail}), which presents a detail of 
the domain near the point G. 
The  area of the removed segment
is $O(L^3/R)$.  The parameter $L$ does not
scale with $R$ when $R$ tends to $\infty$, hence the area 
of the removed segment tends to zero.
Since the volume energy density is finite,
the difference $\tilde E(R)-  E(R) \sim R^{-1}$.

For the BPS saturated configuration the energy $E(R)$
or $\tilde E(R)$ can be rewritten as a combination of the
corresponding contour integrals. The master equation for the wall 
junction is presented in Eq.(\ref{junmasseq}).
{\em It is assumed 
in this equation that 
the phase $\delta$ in Eq.(\ref{eqjunc}) is zero}, which is the case with
our convention. There is an ambiguity in
$\Sigma$ and $a_k$ in this master formula:
a certain reshuffling is possible due to the
ambiguity in the definition of the
supercurrent and the energy-momentum tensor. 
This is explained in great detail
in Ref.\cite{GS}.
We will return to the discussion
of this ambiguity later on. For the time being
let us stick to the {\em canonic} definitions.
Then 
\beq
\Sigma = {\cal W}\,,\quad S_{1,2} = \{{\rm Re}{\cal W}, \, {\rm Im}{\cal W} \}
\,,
\label{dopcurone}
\eeq
and
\beq
 a_k = \frac{1}{2}\sum_\ell \left( \Phi_\ell i\partial_k \bar \Phi_\ell
- \bar \Phi_\ell i\partial_k  \Phi_\ell\right) \equiv 
\frac{1}{2}\sum_\ell \left(\Phi_\ell i 
\stackrel{\leftrightarrow}{\partial_k}\bar\Phi_\ell
\right)\,,
\label{curone}
\eeq
where $k=1,2$. The sum runs over all scalar fields involved in the
solution. Fields which are purely real
obviously drop out. Only complex fields contribute;
for such fields, if $\Phi =\rho e^{i\eta}$,
\beq
a_k = \rho^2 \partial_k\eta\,.
\label{curtwo}
\eeq 
Note that Eq. (\ref{curone}) 
for the current appearing in the master equation
is valid for the wall junctions.
For the anti-junctions one must reverse the sign
in Eqs. (\ref{curone}) and (\ref{curtwo}).

How do we see that with our convention the phase $\delta$ 
in Eq.  (\ref{eqjunc}) is zero? To this end,
let us consider this equation far away from the
``hub'', on the line DH. On this line Eq.  (\ref{eqjunc})
becomes 
\beq
\partial_y\Phi = i \frac{\partial\bar {\cal W}}{\partial\bar\Phi}\,.
\label{eqforw}
\eeq
Since $\Delta{\cal W} = {\cal W}(Q) - {\cal W}(P) = 2i|{\cal W}| \sin(\pi /N)$
in our convention, this is exactly the BPS equation for the (isolated) wall,
as it should be since the equation for the junction should go into the
equation for the wall far away from the ``hub.''

To begin with, we will calculate the tension $T_1$. Since due to the $Z_N$
symmetry, the contributions of each of the $N$
sectors in Fig.(\ref{configuration}) are identical, it is sufficient to 
calculate the integral corresponding to one sector,
\beq
T_1 R = 2N\int_{\rm PDHQ} d\vec{n}_k S_k\,,
\label{horosho}
\eeq
where on PD and HQ the integral runs along the circle while
on DH it runs along the straight line.
On PD and HQ the superpotential ${\cal W}$ assumes the vacuum values
$$
{\cal W} = |{\cal W}|_{\rm vac} e^{\pi i /N}\quad \mbox { on  HQ}\,,\qquad
{\cal W} = |{\cal W}|_{\rm vac} e^{-\pi i /N}\quad \mbox { on  PD}\,,
$$
and does not change (with exponential accuracy).
Therefore,
\beq
\int_{\rm PD+HQ}d\vec{n}_k S_k = R  |{\cal W}|_{\rm vac}\left\{
\int_{-\pi/ N}^{-\gamma}\cos\left(\frac{\pi}{N} +\alpha\right)
+\int_{\gamma}^{\pi /N}\cos\left(\frac{\pi}{N} -\alpha\right)
\right\}\,,
\label{eqstrl2}
\eeq
where the angle $\gamma$ is defined in Fig.(\ref{configuration}).
On the straight line DH we have
\beq
\int_{\rm DH}d\vec{n}_k S_k = \int_{-L/2}^{L/2} d y \, {\rm Re}{\cal W}
\,.
\label{eqstrl}
\eeq
With the exponential accuracy the fields on the straight line DH
satisfy the BPS equations for the isolated wall, see
(\ref{eqforw}). The solution is known to have an integral of motion 
(see e.g. \cite{CS}),
which with our phase convention is nothing but ${\rm Re}\{{\cal W}\}$.
Thus,  ${\rm Re}\{{\cal W}\}$ does not change along the line
DH and reduces to its vacuum value, $|{\cal W}|_{\rm vac}\cos(\pi /N)$,
which is one and the same in both vacua, P and Q. As a result,
the sum of two integrals (\ref{eqstrl2}) and (\ref{eqstrl})
is
\beq
\int_{\rm PDHQ} d\vec{n}_k S_k = R  |{\cal W}|_{\rm vac}\,
2\sin\left(\frac{\pi}{N}
-\gamma\right) +L\, |{\cal W}|_{\rm vac} \cos\frac{\pi}{N}
= 2R|{\cal W}|_{\rm vac}\,
\sin\frac{\pi}{N} +O(R^{-1})\, .
\label{horosh}
\eeq
Moreover,
\beq
2|{\cal W}|_{\rm vac}\,
\sin\frac{\pi}{N}= |{\cal W}(P) - {\cal W}(Q)|
=\frac{1}{2} \,
T_{\rm wall}\, ,
\label{horo}
\eeq
where $T_{\rm wall}$ stands for the tension of the isolated
BPS wall. Combining Eqs. (\ref{horosho}), (\ref{horosh}) and 
(\ref{horo}) we arrive at
\beq
T_1 = NT_{\rm wall}\,,
\eeq
q.e.d.

Now we pass to the discussion of the tension $T_2$.
According to  Eq.(\ref{junmasseq})
for the wall junction tension,
$T_2$ is given by the integral over the large circle over
\beq
T_2 =-\oint  d\vec{x}_k a_k\,,
\label{ploho}
\eeq
where $a_k$ is given in Eq. (\ref{curone}) or Eq. (\ref{curtwo}).
Again, we take into account the fact that all $N$ sectors
give one and the same contribution, and we will
integrate over one sector only, starting from P and ending at Q.
Then
\beq
T_2 =-N\int_{\rm PDHQ}  d\vec{x}_k a_k\, . \label{junten}
\eeq
On the intervals PD and HQ, the fields assume their vacuum values.
They do not change (with exponential accuracy),
and therefore, $a_k =0$, so that there is no contribution
to $T_2$. A  nonvanishing 
contribution comes from the DH segment of the integral.
On this segment we can disregard the ``hub'' and other
``spokes'' of the junction, considering
the relevant part of the wall as that of
the isolated BPS wall. 

In Model I, there is only one field, $X$. In Model II
there are $N+1$ fields, $T$ and $M_\ell$. 
However, for the BPS wall, $M_\ell$'s assume real values
(see Appendix A) and, thus, the fields $M_\ell$
do not contribute. Let us parametrize the fields
$X$ and $T$ as $\rho e^{i\eta}$.
In both cases $\eta (P) = -\pi /N $ and $\eta (Q) = \pi /N $.
Then
\beq
T_2 =-N\int_{\rm DH}  dy\, \rho^2 (x,y)\partial_y \eta (x,y)
= -N \langle\rho^2\rangle\left(\eta (Q)-\eta (P)\right)
= -2\pi \langle\rho^2\rangle\,.
\label{ploh}
\eeq
Here $\langle\rho^2\rangle$ is the average value of the modulus of the field
on the solution on the segment DH, i.e. it is
the average over the wall.
First, we notice that since $\langle\rho^2\rangle$ is positive,
the value of the junction tension $T_{\rm junction}$
is negative. The only way to escape this, is to have some
(complex) fields involved in the junction which rotate in the 
unnatural direction (i.e. while the  ``natural''
fields in the junction
rotate in the anticlock-wise direction,
this ``exotic'' field must rotate clock-wise).
Rotation means here winding of the phase, from smaller to larger values.
In addition, the ``exotic'' field
must have the average value of the modulus larger than
that of the natural fields, so that
it can overcome the negative contribution to $T_{\rm junction}$
coming from the ``natural'' fields.
In all models considered so far of which we are aware,
the ``exotic'' fields do not appear.\footnote{
Let us note in passing that similar consideration determining
the sign of the tension $T_2$ is applicable to the $U(1)$ vortices
of the type considered in \cite{CS}.}

Second, we note that $\langle\rho^2\rangle$
inside the wall does not differ too much from its
vacuum value. For instance, in the Model II
(see Appendix A)
the vacuum value $|T|^2_{\rm vac}= N^2$
while in the middle of the wall 
$|T|_{\rm middle}= N^2[\cos(\pi /N)]^2$.
The average value $\langle\rho^2\rangle$
lies between these two extremes. At large $N$ the average value
tends to its vacuum one, 
$\langle\rho^2\rangle = N^2 + O(1)$.

\vspace{0.2cm}

{\bf Theorem}
For BPS junctions
$T_{\rm junction}$ is determined, through the 
master equation, by the solution for the
isolated wall. Unlike $T_1$,
the domain wall junction is non-holomorphic,
and, generally speaking,  depends on the details of the wall
solution. In the limit $N\to\infty$
the formula for 
$T_{\rm junction}$ becomes universal.

\vspace{0.2cm}

{\bf Proof}

The first part of the theorem is proved by 
the consideration preceding Eq. (\ref{ploh}).
The last part of the theorem will be established below
(see also \cite{GaS}).

\section{$T_{\rm junction}$ from  the isolated wall
in Model I \label{tenI}}

First, we  will illustrate how this works
in Model I.
Let us consider this model in the limit of large $N$.
In this limit the wall junction tension
was calculated in \cite{GS},
$T_{\rm junction}= - 2\pi N^2$.
The corresponding calculation is trivial.
We are now interested in the leading
$1/N$ correction. It can be found
using the solution for the isolated BPS wall
obtained in \cite{DK}.
The field $X$ is parametrized as
\beq
X= N\left(1-\frac{\sigma}{N}
\right)e^{i\tau/N}\,,
\label{GKsol}
\eeq
where
$\sigma (y)$ is a function of $y$
which falls off from
$\sigma_*$ in the middle of the wall ($y=0$) 
to zero at $y=\infty$.
Here $\sigma_*$ is the positive root of the equation
\beq
(1-\sigma_*)e^{\sigma_*} = -1\,,\quad \sigma_* \approx 1.278\,  .
\eeq
The phase $\tau$ changes from zero 
 in the middle of the wall ($y=0$) to $\pi $ at $y=\infty$,
\beq
\tau (\sigma) = \pi - {\rm arccos}\left( (1-\sigma ) \exp\sigma\right)\, ,
\eeq
where we modified the solution presented in 
\cite{DK}
to bring it in accordance with our convention.

Now in the segment under consideration
\beq
a_y =\left(  N - 2\sigma \right) \frac{d\tau}{dy} + O(1/N)\, ,
\eeq
and
$$
- \int_{\rm DH} a_y dy = -2\pi N +4 \int_0^\infty dy \sigma  \frac{d\tau}{dy}
= -2\pi N +4 \int_0^{\sigma_*} \tau (\sigma ) d\sigma 
$$
\beq
= -2\pi N \left( 1-
\frac{1.77}{N} +O(1/N^2)\right)\, .
\eeq
This implies, in turn that
\beq
T_{\rm junction} = - 2\pi N^2
 \left( 1-
\frac{1.77}{N} +O(1/N^2)\right)\, .
\eeq

The tension in Model I can also be calculated numerically for any value
of $N$. In Fig.(\ref{compare}) we compare the numerically determined
tension with the leading and next to leading analytic approximations
for $N=2^k$ and $k=2, ...,10$. The value of $N$ therefore ranges between
$4$ and $1024$. A numerical profile of the domain wall was obtained by
simulation of the second order equation of motion for the field
$X$ on a discrete lattice, using a forward predicting algorithm. 
The lattice spacing was taken much smaller
than the width of the wall, and at the same time the size of the lattice 
was chosen much larger than this width. The value of the field was
fixed to take on the two vacuum values at the two ends of the
one dimensional lattice. A dissipation term was added to the
equation of motion so that the field relaxes to the minimum
energy configuration. An interpolating profile between the
two vacuum expectation values was chosen as the initial condition.
Errors due to the finite lattice spacing are well under control
because the shape of the wall profile is smooth. Errors due to
the finite size of the lattice are suppressed because the wall
profile tends to the vacuum expectation values exponentially fast.
After the field had come to rest, the junction tension was then
obtained from Eq.(\ref{junten}) by numerical integration.

It may be worth a few words to motivate our choice of the second
order equation of motion over the first order BPS
equation to obtain the numerical domain wall profile. Simulation
of the equation of motion yields a domain wall profile even
if the wall is not BPS saturated, but in the case of Model I
this does not offer an advantage, as the walls are known to be
BPS saturated. However, numerical integration of the first order BPS 
equation is fraught with an instability. This can best be
seen from an approximate solution of the BPS equation near a vacuum
value of the field. Parametrizing the field as $X=N(e^{\pi i/N}
+a+bi)$, the solution to the linearized BPS equation for
$a$ and $b$ (valid for $|a|,|b| \ll 1$) is
\begin{eqnarray}
a & = & c_1 \cos\frac{\pi}{N}\, e^{-N^2y} + c_2 \cos\frac{\pi}{N}\,
e^{N^2y} \nonumber \\
b & = & c_1\, (1-\sin\frac{\pi}{N})\, e^{-N^2y} - c_2\, (1+\sin\frac{\pi}{N})\,
e^{N^2y}\, ,
\end{eqnarray}
where $c_1$ and $c_2$ are real constants.
For a wall connecting the vacuum with $k=1$ at $y=-\infty$ to
the vacuum with $k=2$ at $y=\infty$, $c_2=0$. The value
of $c_1$ is associated with the location of the center of the
wall. In simple numerical procedures to integrate 
the BPS equation, $c_2$ will be small, but not equal to zero.
As a result, the numerical solution will show runaway behavior
far away from the center of the wall.
\begin{figure}
\epsfxsize=10cm
\centerline{\epsfbox{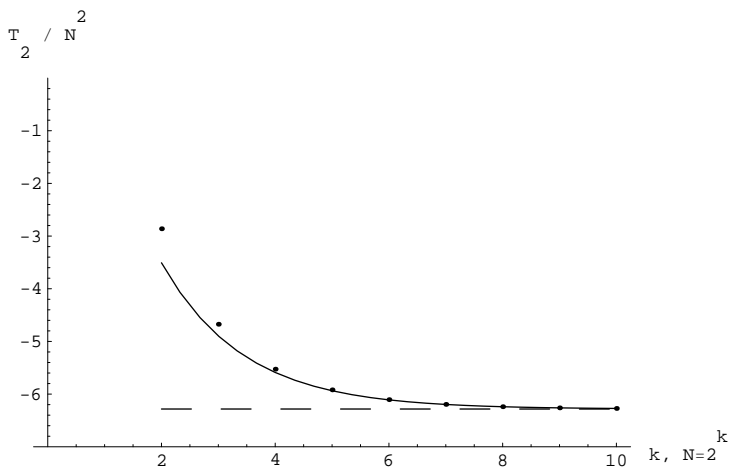}}
\caption{The tension of the junction in Model I as a function
of $N$. The dashed line reflects the leading order contribution in
$1/N$. The solid line includes the next to leading order contribution, and
the dots represent the numerical calculation.}
\label{compare}
\end{figure}

\vspace{0.2cm}

\section{$T_{\rm junction}$ in Model II \label{tensII}}

Although in general $T_{\rm junction}$ depends on the detailed
shape of the domain walls surrounding the junction, this
is not the case for Model II.
In fact, the tension of the junction in Model II does {\em not}
depend on $A$ and $B$ and can be calculated exactly even
though an analytic solution for the domain wall is not available
for generic ratios of $A$ and $B$.
The only assumption is that the fields $M_1$ and $M_2$ are real on the domain
wall profile. This is consistent with the explicit domain wall solution
for the special ratio of $A$ and
$B$ discussed in Appendix A, and it is confirmed by numerical solutions
of the domain wall profile for generic ratios.
The field $T$ then takes the form $T=-i t + N \cos\pi/N$ on the domain
wall, where $t$ is a real function with asymptotic values 
$t(-\infty)=N\sin\pi/N$ and $t(\infty)=-N\sin\pi/N$.
The junction tension is now equal to
\begin{equation}
T_2 = - \oint dx_k a_k = 
N \int_{-\infty}^{\infty} N \cos\pi/N \frac{dt}{dx} dx
=-N^3 \sin 2\pi/N. \label{tenII}
\end{equation} 
The tension therefore only depends on the asymptotic values
of the field $T$ on the domain wall and is independent
of the detailed shape of the wall. In this sense, the 
 junction tension $T_{\rm junction}$  in Model II is similar to the $(1,0)$
tension of domain walls.

\section{Reshuffling between the $(1,0)$ and $(1/2,1/2)$
central charges and the unique expression for $T_{1,2}$ \label{ambiguity}}

As was discussed at length in \cite{GS},
in the case of wall junctions 
in supersymmetric theories, one deals, in fact, with
two central charges -- one appears in the anticommutator
$\{Q_\alpha Q_\beta\}$ (this is the 
$(1,0)$ central charge), another appears in the
anticommutator
$\{Q_\alpha\bar{Q}_{\dot\beta}\}$ (this is the 
$(1/2,1/2)$ central charge).
The individual  expressions for these central charges
are not unique. They depend on the expression for the
supercurrent one starts with. This is 
due to the fact
that the effect under discussion is subtle --
we are interested in integrals over total derivatives.
The supercurrent has an ambiguity. Say,
in the Wess-Zumino model with one chiral  superfield
one can add to the supercurrent the term
\beq
\Delta J_{\alpha\beta\dot\beta}
=-\frac{\sqrt{2}}{3}
\left[\partial_{\alpha\dot\beta}\left(\psi_\beta\phi^\dagger
\right)+\partial_{\beta\dot\beta}\left(\psi_\alpha\phi^\dagger
\right)
-3\varepsilon_{\beta\alpha}\partial^\gamma_{\dot\beta}
\left(
\psi_\gamma\phi^\dagger\right)
\right]\,,
\label{dopone}
\eeq
which is conserved (nondynamically) and presents a full derivative.
It has no impact on the supercharges
defined as
\beq
Q_\alpha =\int d^3 x J^0_\alpha\,,\quad
J^\mu_\alpha =\frac{1}{2}\left(\bar{\sigma}^\mu
\right)^{\dot\beta\beta} J_{\alpha\beta\dot\beta}
\eeq
(for definitions see \cite{CS}).
The contribution of $\Delta J_{\alpha\beta\dot\beta}$
in $Q_\alpha$ is of the form $\int d^3x \vec\nabla (...)$.
Adding the term in Eq. (\ref{dopone})
one changes both central charges.

If we do not add the term $\Delta J_{\alpha\beta\dot\beta}$
in the supercurrent at all (by the way, this will correspond to the
canonic energy-momentum tensor), then we get Eq.(\ref{junmasseq})
with $S$ and $a$ defined in Eqs. (\ref{dopcurone}) and (\ref{curone}).
Assume that we add to the supercurrent the term 
$\Delta J_{\alpha\beta\dot\beta}$ as in Eq. (\ref{dopone}),
with the coefficient indicated in this expression. What changes?

It is not difficult to show  \cite{CS}
that the changes are as follows:
(ii) the energy-momentum tensor appearing in the
anticommutator of two supercharges acquires
a full-derivative (conserved) term built from the scalar fields;
this term is such that $\theta^\mu_\mu$ now vanishes for
purely cubic superpotential, i.e. for the conformally invariant theory;
(ii) $\Sigma$ in Eq.  (\ref{dopcurone})
now becomes
\beq
\Sigma ={\cal W} - \frac{1}{3}\Phi\frac{\partial{\cal W}}{\partial\Phi}\,,
\quad S_{1,2} =\{{\rm Re} \Sigma\,,\,{\rm Im}\Sigma \} ;
\label{newsup}
\eeq
(iii) simultaneously, the expression
$-\oint a_k dx_k$ in Eq.(\ref{junmasseq})
changes too. Let us keep, for simplicity, the definition of the
current (\ref{curone}) intact.
Then a ``new'' master equation takes the form
\beq
\frac{M}{{\rm length}}= -\frac{1}{3}\oint a_kdx_k
+2\oint dn_k S_k\,,
\eeq 
where $S$ is defined in Eq.(\ref{newsup}).

It is instructive to check that, in spite of the reshuffling
in the $(1,0)$ and $(1/2,1/2)$ central charges,
the final results  for the tensions $T_{1,2}$
remain unambiguous. They are insensitive to this reshuffling.
One can do this check explicitly using the analytic solution for the
BPS wall which exist in Model II, see Appendix. 
(A general proof was presented in \cite{GS}.)
As previously, we split the contour integral in $N$ sectors;
we will explicitly consider only the transition PQ (Fig.(\ref{configuration})).
On the segments PD and HQ the extra term in the new master
equation
$\sum \Phi \partial{\cal W}/\partial\Phi$ vanishes, since 
these segments correspond to vacua, where $ \partial{\cal W}/\partial\Phi =0$.
It is only the interval DH that contributes
to the integral over $(-2/3)\sum \Phi \partial{\cal W}/\partial\Phi$.
One can readily convince oneself that
the contribution due to $M_{1,2}$ appears only in Im$\Sigma \equiv S_2$,
and, thus, drops out. Only the field $T$ is relevant.
Due to the fact that only the segments of the type
DH contribute, it is immediately clear
that the above extra term in the new master equation
changes only the tension $T_2$ leaving $T_1$ intact.

Now, at $R\to \infty$ one can replace the DH segment by that
for an isolated BPS wall, 
using explicit expressions collected in Appendix A.
In this way we get
\beq
-\frac{2}{3}
\int \, dn_1\,\mbox{Re}\left\{ \Phi\frac{ \partial{\cal W}}{\partial\Phi}\right\}
=-\frac{2}{3}N^2 \sin \frac{2\pi}{N} \, ,
\label{netime}
\eeq
and, consequently,
\beq
{2}
\oint \,dn_k\Delta S_k =-\frac{2}{3}N^3 \sin \frac{2\pi}{N}
\equiv \delta_2(T_2)\, ,
\label{netimee}
\eeq
where 
\beq
\Delta S_{1,2} = \left\{{\rm Re}\left(-\frac{1}{3}\Phi
 \frac{\partial{\cal W}}{\partial\Phi} \right)\,,\,  {\rm Im}
\left(-\frac{1}{3}\Phi
 \frac{\partial{\cal W}}{\partial\Phi} \right)
\right\}\,.
\eeq
Using the same approach we readily find that
the term
\beq
T_2 = - \oint dx_k a_k =-N^3\sin \frac{2\pi}{N}\, ;
\label{nettwo}
\eeq
this term was in the old master equation.
In the new master equation we have one third of it, i.e.
the contribution to $T_2$ coming from the axial current is
one third of that in Eq. (\ref{nettwo}),
\beq
\delta_1 (T_2)  = -\frac{1}{3} N^3\sin \frac{2\pi}{N}\, .
\label{deltone}
\eeq
Combining this with $\delta_2(T_2)$ from Eq. (\ref{netimee}),
we arrive at 
\beq
T_2 = \delta_1(T_2) +\delta_2(T_2) =-N^3\sin \frac{2\pi}{N}\, ,
\label{tensionII}
\eeq
exactly the same result for $T_{\rm junction}$
that was obtained from the old master equation where only the
axial current contributed to it.
The reshuffling in the central charges does take place;
there is no impact on the physical quantities, however.

\section{BPS saturation of junctions in Models I and II \label{saturation}}

In this section we consider the question whether the domain wall
junctions in Models I and II saturate the BPS bound.
For certain limiting values of the parameters, the BPS equations for 
domain wall junctions in Models I and II allow an analytic solution. 
This is the case for the $N \rightarrow \infty$ limit
in model I, as was shown in \cite{GS}. And in Appendix B we present 
an analytic solution for the triple junction in Model II, which
can be obtained when $N$ is a multiple of $3$ and for a particular
ratio of $A$ and $B$. This junction is a generalization of the
analytic solution in \cite{OINS}. For $N=3$ the triple junction
is also the basic junction. The above values of the parameters
seem only special because they allow an analytical solution, and
not for any other reason. Moreover, for any value of the parameters
there are sensible approximate solutions to the BPS equation
far away from the center of the junction (domain walls surrounding
the center) and near the center (string-like solutions). All
this information strongly suggests that the junctions are BPS
saturated for all values of the parameters. 
 
In order to further test this hypothesis, we performed
a numerical analysis. We simulated the second order equations
of motion for both Models I and II on a lattice using a simple
forward predicting algorithm. The procedure
is a two-dimensional generalization of the procedure described in 
Sect.(\ref{tenI}) to generate numerical domain wall profiles. The 
lattice spacing
was chosen to be much smaller than the characteristic size $l$
of the junctions, and at the same time the size of the lattice
was much larger than $l$. We put the appropriate domain wall profiles
(numerically generated)
on the edges of the lattice as boundary conditions. A dissipation
term was added to the equations of motion, so that the field
configuration relaxes to the configuration with lowest energy
consistent with the boundary conditions. 

We then determined whether the numerically obtained junction profile
saturated the BPS bound by comparing the energy of the configuration to
$E(R)=T_1 R + T_2$. For Model I, $T_1=4 N^4/(N+1) \sin\pi/N$, and $T_2$
was determined numerically as in Sect.(\ref{tenI}). For Model
II, $T_1=4 N^3 B \sin \pi/N$ and $T_2$ is given in Eq.(\ref{tenII}).
Numerical errors due to the finite lattice spacing were well under
control because the junction shape is smooth, and errors
due to the finite size of the lattice were suppressed because
the junction profile tends toward the domain wall solutions exponentially
fast away from its center.

We performed this procedure for Model I with $N=4$, and for
Model II with $N=4$ and various values of the parameters $A$
and $B$. In each case we found that the junction was
indeed BPS saturated. This fact, in conjunction with the 
analytical results, lends strong support to the conjecture
that the basic junctions we have discussed in Models I and II
are BPS saturated for all values of the parameters.

\section{Comments on Stability of the BPS wall junctions \label{stability}}

A general question which deserves a brief discussion
is the stability (local and global) of the BPS wall junctions.
In  this issue two features are crucial:
(i) the two-dimensional manifold perpendicular
to the junction axis (the plane) is noncompact,
i.e. the junction ``spokes'' extend to infinity;
(ii) the wall junction is {\em not} a two-dimensional soliton;
rather,
each ``spoke'' represents a wall 
which extends in the direction perpendicular to the
plane. This extension is infinite.

As was discussed above,
the mass of the 
BPS wall junction can be expressed in terms of two central
charges which reduce, in turn,
to the contour integral over the large circle
(see Fig.(\ref{fourfig}.a), where we consider 
a $Z_6$ configuration, for definiteness).
This means that any {\em localized}
perturbation of the
junction configuration,
both in the plane (Fig.\ref{fourfig}.b), and in the perpendicular direction,
will lead to an encreased energy, i.e.
the BPS wall junction
is stable against such localized perturbations.

A global perturbation, when two ``spokes''
approach each other and then eventually
fuse into one, is energetically favorable
(Figs. (\ref{fourfig}.c) and (\ref{fourfig}.d). This is due to the fact
that the isolated BPS wall tension
for the transition from the vacuum I to the vacuum J
has the form 
\beq
T_{\rm IJ} = 2|{\cal W}_{\rm J} - {\cal W}_{\rm I}|
= 2|{\cal W}_{\rm vac}|\times |e^{2\pi i J /N} - e^{2\pi i I /N}|\, ,
\eeq
in the models under consideration, and, hence
\beq
T_{\rm I-II} + T_{\rm II-III} > T_{\rm I-III}\, .
\eeq
In other words, eating up an intermediate vacuum
we lower the energy of the configuration.

Were the wall sizes finite, the configuration
with six spokes in Fig.(\ref{fourfig}.a) could quantum-mechanically
tunnel into the configuration with five spokes (Fig.(\ref{fourfig}.d),
under the barrier. However,
since the walls are infinite in both dimensions,
this cannot happen.
The boundary conditions at infinities make the
BPS wall junctions stable both classically
and quantum-mechanically. For further consideration see Ref. \cite{BT}.

\begin{figure}
\epsfxsize=9cm
\centerline{\epsfbox{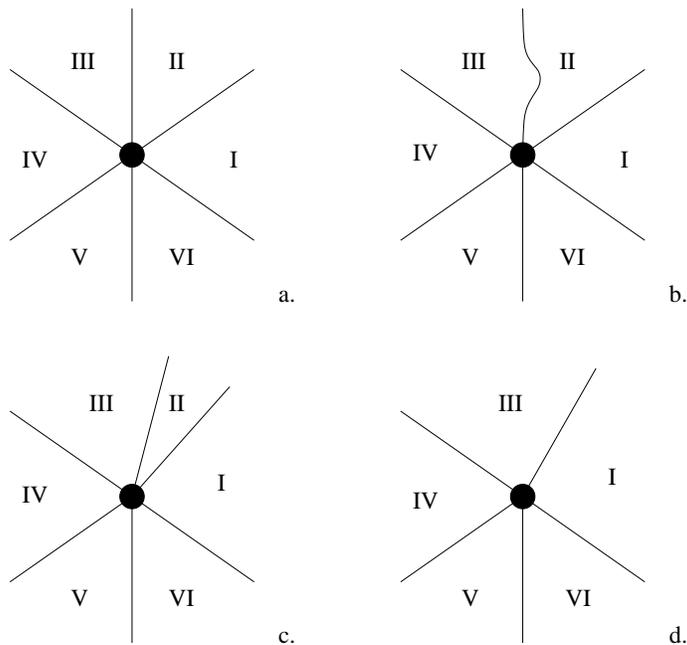}}
\caption{Stability of the wall junction. a.) Regular BPS junction in a model 
with $Z_6$. b.) A localized deformation 
of the wall junction. c.) Squeezing out the vacuum II.
d.) The wall junction with the vacuum II missing.}
\label{fourfig}
\end{figure}

\section{Conclusions}
In supersymmetric models with the spontaneously broken $Z_N$
 symmetry we developed techniques 
for (i) establishing the BPS nature of the wall junctions of the 
``hub and spokes" type and for  (ii) calculating the tensions
$T_1$ and $T_{\rm junction}$. 
It is shown that in ``natural" models $T_{\rm junction}<0$.
The conditions under which the ``unnatural" sign $T_{\rm junction}>0$
can occur are discussed. 

\section*{Acknowledgements}
{
One of the authors (M.S.) acknowledges useful discussions
with Zurab Kakushadze which took place at ITP, Santa Barbara,
in the course of the program ``Supersymmetric Gauge Dynamics and String Theory".
This work was
supported in part by the Department  of Energy under Grant No.
DE-FG02-94ER40823 and by National Science
Foundation under Grant No. PHY94-07194.}

\section*{Appendix A}

Here we present an analytic solution for the BPS wall in 
Model II
which can be readily found
for a specific ratio of the coefficients $A$
and $B$, namely
\beq
A = \frac{B}{\left[2N\sin (\pi /N) \right]^2}\,.
\label{luch}
\eeq
For what follows
it is convenient to introduce the parameter
$\lambda$ which determines the width of the wall,
\beq
\lambda = l^{-1} = \frac{B}{2\sin (\pi /N)}\,.
\label{luch2}
\eeq
Note that at large $N$ the parameter $\lambda$
scales as $N$; correspondingly, $l\sim N^{-1}$.

We will consider the wall interpolating
between the vacuum 1 (``initial'' vacuum)
at $y=-\infty$ and the vacuum 2 (``final'' vacuum)
at $y = +\infty$.

The vacuum 1 lies at
\beq
T= Ne^{-\pi i /N}\,,\quad M_1 = \sqrt{\frac{B}{A}}\,,
\quad M_2 =M_3=.... = 0\, .
\label{luch3}
\eeq
The vacuum 2 lies at
\beq
T= Ne^{\pi i /N}\,,\quad M_2 = \sqrt{\frac{B}{A}}\,,
\quad M_1 =M_3=.... = 0\, .
\label{luch4}
\eeq
For the superpotential
in the vacua 1 and 2 we have
\beq
{\cal W}(1) = BN^2e^{-\pi i /N}\,,\quad {\cal W}(2) = BN^2e^{\pi i /N}\,,
\label{luch5}
\eeq
so that
\beq
\Delta {\cal W} = 2 iBN^2\sin(\pi  /N)\,.
\eeq
Correspondingly, the BPS equations for the fields $T$, $M_{1,2}$
have the form
\beq
\dot T = i \frac{\partial\bar{\cal W}}{\partial \bar T}\, ,\quad
\dot M_\ell = i \frac{\partial\bar{\cal W}}{\partial \bar M_\ell}\,.
\label{luch6}
\eeq
Here the dot denotes differentiation over $y$.
The solution of these equations is
\beq
T = N\cos\frac{\pi}{N} + iN\left(\sin\frac{\pi}{N}\right)
{\rm tanh} (\lambda y )\, ,
\label{luch7}
\eeq
and
\beq
M_1 =\sqrt{\frac{B}{A}}\, \frac{1-{\rm tanh} (\lambda y ) }{2}\,,
\quad
M_2 =\sqrt{\frac{B}{A}}\, \frac{1+{\rm tanh} (\lambda y ) }{2}\,.
\label{luch8}
\eeq

\section*{Appendix B}

Here we present an analytic solution for BPS walls and
triple junctions in 
Model II, with $N$ a multiple of $3$.
An analytic solution for the triple junction exists
if the coefficients $A$
and $B$ satisfy the relation
\beq
A = \frac{B}{3 N^2}\,.
\label{}
\eeq
Note that the ratio of $A$ and $B$ depends on $N$, in contrast
with what was assumed in the main text.
For this particular ratio of $A$ and $B$, the parameter
$\lambda$ which determines the width of the wall and the
size of the junction takes the form
\beq
\lambda = l^{-1} = \frac{B}{\sqrt{3}}\,.
\label{scale}
\eeq
We will consider a junction that interpolates
between three of the $N$ vacua.
Vacuum 1 lies at
\beq
T= N e^{-\pi i /N}\,,\quad M_1 = \sqrt{\frac{B}{A}}\,,
\quad M_k= 0\, ,k \neq 1,
\label{vac1}
\eeq
vacuum 2 lies at
\beq
T= N e^{2 \pi i /3} e^{-\pi i /N}
\,,\quad M_{N/3+1} = \sqrt{\frac{B}{A}}\,,
\quad M_k= 0\,,k \neq N/3+1,
\label{vac2}
\eeq
and vacuum 3 lies at
\beq
T= N e^{4 \pi i /3} e^{- \pi i /N}
\,,\quad M_{2N/3+1} = \sqrt{\frac{B}{A}}\,,
\quad M_k=0\, ,k \neq 2N/3+1.
\label{vac3}
\eeq
Let us first focus on a 
domain wall interpolating between vacua 1  and 2 along the
$\hat{y}$ axis. Such a wall is a solution to the 
BPS equations 
\beq
\dot{T}= i e^{i \delta} \frac{\partial\bar{\cal W}}
{\partial \bar T}\, ,\quad
\dot{M}_1 = i e^{i \delta} \frac{\partial\bar{\cal W}}
{\partial \bar M_1}\, ,\quad
\dot{M}_{N/3+1} = i e^{i \delta} \frac{\partial\bar{\cal W}}
{\partial \bar M_{N/3+1}}\, ,
\label{bpswall}
\eeq
where the dot indicates differentiation with respect to $y$.
When the phase takes the value $\delta= \pi/3-\pi/N$, the wall 
configuration tends to vacuum 1 at $y\rightarrow-\infty$
and to vacuum 2 at $y\rightarrow \infty$. The analytic solution
to the BPS equation is
\beq
T = \frac{N}{2} e^{-\pi i/N} \left( e^{\pi i/3} -
\sqrt{3} e^{-\pi i/6} 
{\rm tanh} (\lambda y )\right)\, ,
\label{Twall}
\eeq
and
\beq
M_1 =\sqrt{\frac{B}{A}}\, \frac{1-{\rm tanh} (\lambda y ) }{2}\,,
\quad
M_{N/3+1} =\sqrt{\frac{B}{A}}\, \frac{1+{\rm tanh} (\lambda y ) }{2}\,.
\label{Mwall}
\eeq
Having the solution for the wall in hand, we now continue with
the triple junction. In accordance with the convention in the
main text, our choice of $\delta$ implies that the spoke
in between the first and the second vacuum points in the
direction of the positive $\hat{x}$ axis. For $N=3$,  
the phase $\delta=0$, as in the main text. For
other values of $N$, insisting on $\delta=0$ would lead to a very
awkward orientation of the junction. The junction is the solution to 
the BPS equations
\beq
2 \frac{\partial T}{\partial \zeta} = e^{i \delta} \frac{\partial\bar{\cal W}}
{\partial \bar T}\, \nonumber
\eeq
and
\beq
2 \frac{\partial M_1}{\partial \zeta} = e^{i \delta} \frac{\partial\bar{\cal W}}
{\partial \bar M_1}\, ,\quad
2 \frac{\partial M_{N/3+1}}{\partial \zeta} = e^{i \delta} \frac{\partial\bar{\cal W}}
{\partial \bar M_{N/3+1}}\, ,\quad
2 \frac{\partial M_{2N/3+1}}{\partial \zeta} = e^{i \delta} \frac{\partial\bar{\cal W}}
{\partial \bar M_{2N/3+1}}\, .
\label{bpsjunction}
\eeq
with $\zeta =x+i y$. Defining
\beq
g(x,y) = e^{-\lambda(y-x/\sqrt{3})} + e^{\lambda(y+x/\sqrt{3})}
+ e^{-2\lambda x /\sqrt{3}},
\eeq
the triple junction BPS configuration takes the form 
\beq
T = N e^{-\pi i/N} \frac{
e^{-\lambda(y-x/\sqrt{3})} + e^{2 \pi i/3} e^{\lambda(y+x/\sqrt{3})}
+ e^{4 \pi i/3} e^{-2\lambda x /\sqrt{3}} }{g(x,y)},
\eeq
and
\beq
M_1 = \sqrt{\frac{B}{A}} \frac{e^{-\lambda(y-x/\sqrt{3})}}{g(x,y)}\, ,\quad
M_{N/3+1} = \sqrt{\frac{B}{A}} \frac{e^{\lambda(y+x/\sqrt{3})}}{g(x,y)}\, ,\quad
M_{2N/3+1} = \sqrt{\frac{B}{A}} \frac{e^{- 2 \lambda x/\sqrt{3}}}{g(x,y)}\, .
\eeq
For $x\rightarrow \infty$ the solution tends to the domain wall 
configuration along the $\hat{y}$ axis which was discussed before. 
For $x\rightarrow
-\infty$  and $y=0$ the field configuration tends towards vacuum 3.
The tension for the triple junction is
\beq
T_{2} = - \frac{3}{2} \sqrt{3} N^2.
\label{tension}
\eeq
This is an exact result, valid for any value of $A$ and $B$, 
not just the special ratio
for which the analytic solution for the junction is obtained.
For $N=3$ this result agrees with the tension presented in 
Eq.(\ref{tensionII}) in the main text.

\end{document}